# GPCALMA, a mammographic CAD in a GRID connection

U. Bottigli[1], P. Cerello[2], P. Delogu[3], M.E. Fantacci[3], F. Fauci[4], B. Golosio[1], A. Lauria[5], E. Lopez Torres[2], R. Magro[4], G.L. Masala[1]♦, P. Oliva[1], R. Palmiero[5], G. Raso[4], A. Retico[3], S. Stumbo[1], S. Tangaro[6]

[1]Struttura Dipartimentale di Matematica e Fisica dell'Università di Sassari and Sezione INFN di Cagliari, via Vienna 2, Sassari 07100 Italy; [2]Sezione INFN di Torino via Giuria 1,10125 Torino, Italy; [3]Dipartimento di Fisica dell'Università and Sezione INFN di Pisa, Via Buonarroti 2,56100 Pisa, Italy [4]Dipartimento di Fisica e Tecnologie Relative dell'Università di Palermo and Sezione INFN di Catania, viale delle Scienze (Parco d'Orleans) 90128 Palermo,Italy; [5]Dipartimento di Scienze Fisiche dell'Università "Federico II" and Sezione INFN di Napoli, Complesso Universitario Monte S. Angelo, Via Cinthia, I-80126, Napoli, Italy; [6]Dipartimento di Fisica dell'Università di Bari and Sezione INFN di Cagliari, via Amendolia 173, 70126 Bari Italy; ♦ Corresponding author: giovanni.masala@ca.infn.it , Ph.: +39 079229486 Fax: +39079229482

**Abstract**

Purpose of this work is the development of an automatic system which could be useful for radiologists in the investigation of breast cancer. A breast neoplasia is often marked by the presence of microcalcifications and massive lesions in the mammogram: hence the need for tools able to recognize such lesions at an early stage. GPCALMA (Grid Platform Computer Assisted Library for MAmmography), a collaboration among italian physicists and radiologists, has built a large distributed database of digitized mammographic images (at this moment about 5500 images corresponding to 1650 patients). This collaboration has developed a CAD (Computer Aided Detection) system which, installed in an integrated station, can also be used for digitization, as archive and to perform statistical analysis. With a GRID configuration it would be possible for the clinicians tele- and co-working in new and innovative groupings ('virtual organisations') and, using the whole database, by the GPCALMA tools several analysis can be performed. Furthermore the GPCALMA system allows to be abreast of the CAD technical progressing into several hospital locations always with remote working by GRID connection. We report in this work the results obtained by the GPCALMA CAD software implemented with a GRID connection.

**Introducion**

Breast cancer is reported as one of the first causes of women mortality [1] and an early diagnosis of breast cancer in asymptomatic women makes it possible the reduction of breast cancer mortality: in spite of a growing number of detected cancers, the death rate for this pathology decreased in the last 10 years [2], thanks also to early diagnosis, made possible by screening programs [3]. Presently, an early diagnosis is possible thanks to screening programs, which consist in a mammographic examination performed for 49-69 years old women. Mammography is widely recognized as the only imaging modality for the early detection of the abnormalities which indicate the presence of a breast cancer [4]; it is realized by screen-film modality or, more recently, by digital detectors [5]. It has been estimated that radiologists involved in screening programs fail to detect up to approximately 25% breast cancers visible on retrospective reviews and that this percentage increases if minimal signs are considered [6]. Sensitivity (percentage of pathologic images correctly classified) and specificity (percentage of non pathologic images correctly classified) of this examination increase if the images are independentlyanalysed by two radiologists [7]. So independent double reading is now strongly recommended as it allows to reduce the rate of false negative examinations by 5-15% [8]. The aim of the GPCALMA collaboration is the behaviour as a "second reader". We present here the status of the art of this project: a large database of digitized mammograms, CAD programs for masses and microcalcification detection, an integrated station in a GRID connection. Furthermore stations have been implemented and are currently on clinical trial in some italian hospitals.

**Methods**

The CAD system here presented is composed by an expert system based on a neural network classifier and a pleasant graphics users interface (GUI). The GUI, by means of a facility tool for the image visualisation and elaboration, provides the support for medical diagnosis on a high-resolution screen.

The images (18x24 $cm^2$, digitized by a CCD linear scanner with a 85 μm pitch and 4096 grey levels) are fully characterised: pathological ones have a consistent description which includes radiological diagnosis and histological data, while non pathological ones correspond to patients with a follow up of at least three years [9].

The automated microcalcification clusters analysis, i.e. the search of groups of small and brilliant objects of different shape and intensity in a very noisy background, is made using a hybrid approach containing both algorithms and neural networks, by which the ROIs (Regions Of Interest) are extracted. A microcalcification is a rather small (0.1 to 1.0 mm in diameter) but very brilliant object. Some of them, either grouped in clusters or isolated, may indicate the presence of a tumour. In our database, the average diameter of microcalcification clusters, as indicated by our radiologists, is 2.3 cm.

The microcalcification cluster analysis was made using the following approach:
- the digital mammogram is divided into 60x60 pixels wide windows;
- the windows outside the breast image are rejected;
- the windows are statistically selected comparing the local and the global maxima;
- the windows are shrunk from 60x60 to 7x7 and are classified (with or without microcalcification clusters) using a FFNN with 49 input, 6 hidden, and 2 output neurons;

- the windows are processed by a convolution filter to reduce the large structures;
- a self-organizing map (a Sanger's neural network) analyzes each window and produces 8 principal components;
- the principal components are used as input of a FFNN able to classify the windows matched to a threshold (the response of the output neuron of the neural network);
- the windows are sorted by the threshold;
- at most three windows are memorized, if their associated probability exceeds a given threshold;
- the selected windows are zoomed to 180x180 pixels, i.e. 1.5x1.5 cm$^2$;
- the overlapping windows are clusterized.

The ROI are indicated on the images and a probability of containing a microcalcification cluster is associated to each ROI. The ROI are shown in the figure 1.

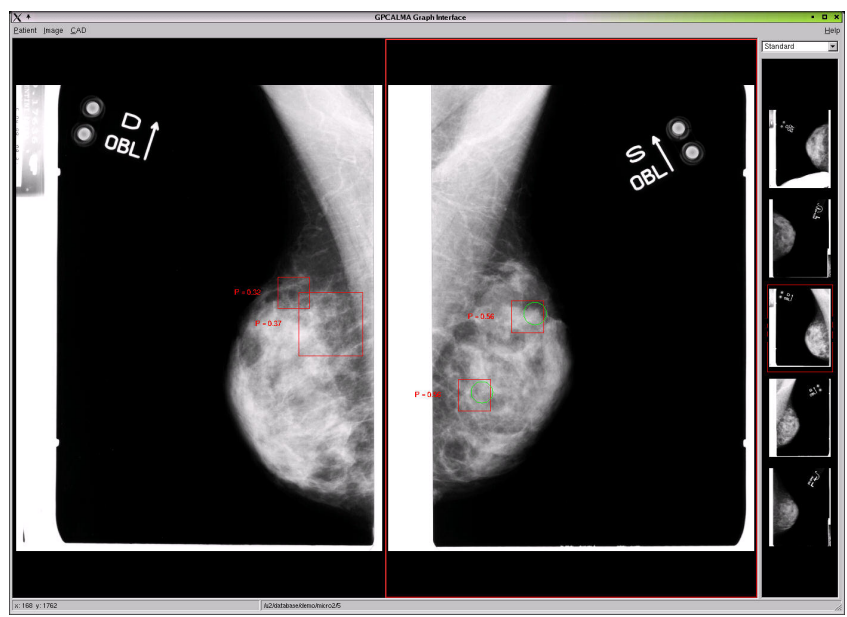

Fig 1: In the image, the CAD application is shown. On the right the green circle rappresent medical correct diagnosis; the red squares rappresent the microcalcification cluster (with probability) found by CAD. On the left image we have two false-positive with low probability.

The automated massive lesions analysis, i.e. the search for rather 'large objects' usually characterized by peculiar shapes, is also made using neural network, but by different algorithms of features extraction and with a different architecture. Massive cluster analysis, shown in figure 3, was made using the following approach:

- Select maximum intensity position, by starting from left top corner of the mammogram the absolute maximum of intensity is found

- A set of concentric rings, 5 pixels wide, up to a maximum radius of 247 pixels (~ 2 cm) is built.
- The pixel average intensity in each ring is computed.
- The most external ring, which defines the ROI radius (R), is that ring whose average intensity is less than a given minimum threshold.
- The entire portion inside the ROI radius is then removed and stored for further analysis.
- A new maximum (a new centre) is sought in the remaining matrix.
- Back to the beginning until one of the following condition is verified :
    a)   100 maximum are found
    b)   the n-th maximum intensity is less than a threshold
- For radius, r = R, 2/3 R, 1/3 R , the parameters are used as INPUT in a FFNN with 9 input, 6 hidden, and 1 output neurons to distinguish between pathological and non-pathological ROI.

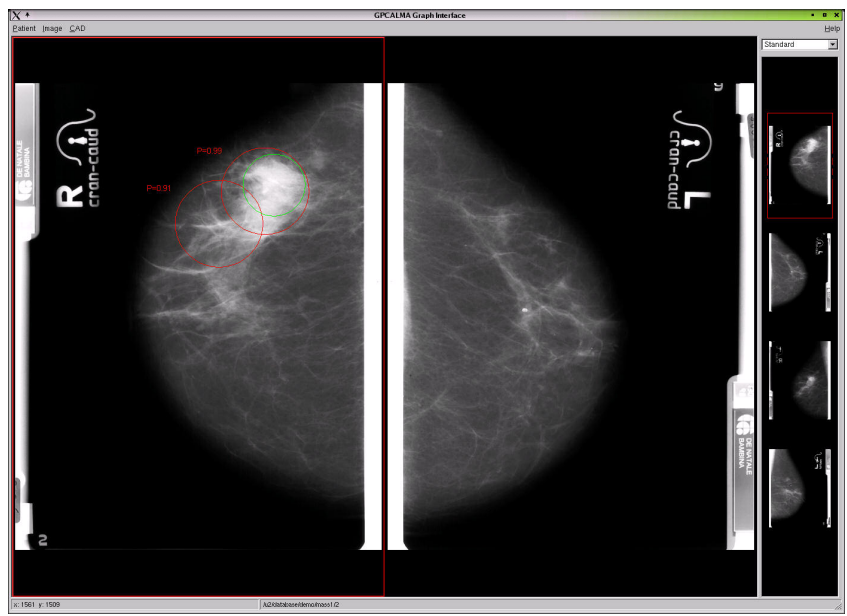

Fig 2: In the left image,the CAD application is shown. The green circle rappresent medical correct diagnosis; the red circles rappresent the massive lesion found by CAD.

The distributed database will be obtained by connecting the hospitals and research centres using GRID technology. In each hospital the images are stored in a local database. Using GRID connection, GPCALMA allows the work with the shared database data as well as with the local database data. Each image, not compressed, fits to 10.5 Mbytes, therefore the transfer of all patient images from a local database to another place could require a large amount of time, so the images transfer is not suitable. The data of each patient are shown to the user as if they were stored into a large central database alone rather than into a shared database. This is possible by using the metadata abstracting information. The system performance are optimized for reducing the time delay when the user is working with the remote images. The GRID approach to mammographic CAD is based on the assumption that it's better to move code rather data. Therefore the user has the advantage to work using the CAD on the remote patient images without the transfer of the images and without knowing their real location. This approach preserves data patient integrity and provides the opportunity for the statistical studies on geographical, cultural, environmental and temporal influences on breast cancer.

**Results**

Using the parameters sensitivity (percentage of pathologic images correctly classified) and specificity (percentage of non pathologic images correctly classified) , the results obtained with this analysis are described in terms of the ROC (Receiver Operating Characteristic) curve, which shows the true positive fraction (sensitivity), as a function of the false positive fraction  (1-specificity) obtained varying the threshold level of the ROI selection procedure. In this way, the ROC curve produced by GPCALMA stations allows the radiologist to detect microcalcification clusters and massive lesions with predictable performance, so as he can set the desired true-positives fraction value and know the corresponding false-positives fraction value: all that will allow to plan the mammographic trials and to use this tool on the basis of local requirements. The ROC curve are shown in figure 3,4 for massive lesions and microcalcification clusters, respectively. The best results for the microcalcification clusters analysis are 92% for both sensitivity and specificity. Instead the best results of the automated massive lesions analysis are 94% for sensivity and 95% for specificity.

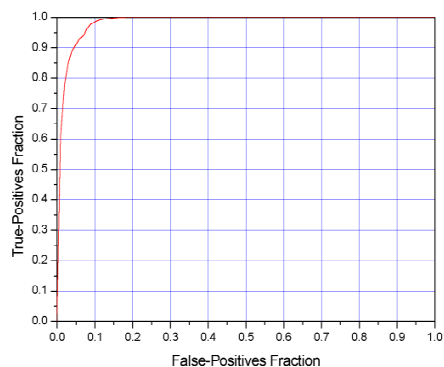

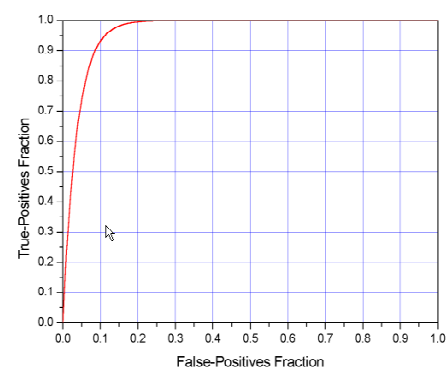

Figure 3, ROC curve for massive lesions 514 mammograms (102 *with* 412*without* massive lesions)

Figure 4, ROC curve for microcalcifications detection; 865 mammograms (370 *with* 495 *without* microcalcifications )

**Discussion**

The results obtained by the GPCALMA Collaboration in the classification of microcalcifications clusters and massive lesions have been compared with commercial CAD systems and human performance [10] and the GPCALMA CAD has been tested as a "second reader" in the classification of the breast disease.

GPCALMA compares very well with other commercial CAD systems and the results obtained in a clinical trial performed using the GPCALMA CAD as a "second reader" show that its use accounts for a substantial increase in sensitivity with an associated small decrease in specificity, and that the numerical values of these depend on the experience of the human "first" reader.

In this moment we are making some test for the GRID connection and a server is installed in the INFN center in Turin for the virtual organization management.